
\input phyzzx
\catcode`\@=11
\paperfootline={\hss\iffrontpage\else\ifp@genum\tenrm
 -- \folio\ --\hss\fi\fi}
\def\titlestyle#1{\par\begingroup \titleparagraphs
 \iftwelv@\fourteenpoint\fourteenbf\else\twelvepoint\twelvebf\fi
 \noindent #1\par\endgroup }
\def\GENITEM#1;#2{\par \hangafter=0 \hangindent=#1
 \Textindent{#2}\ignorespaces}
\def\address#1{\par\kern 5pt\titlestyle{\twelvepoint\sl #1}}
\def\abstract{\par\dimen@=\prevdepth \hrule height\z@ \prevdepth=\dimen@
 \vskip\frontpageskip\centerline{\fourteencp Abstract}\vskip\headskip }
\newif\ifYUKAWA  \YUKAWAtrue
\font\elevenmib   =cmmib10 scaled\magstephalf   \skewchar\elevenmib='177
\def\YUKAWAmark{\hbox{\elevenmib
 Yukawa\hskip0.05cm Institute\hskip0.05cm Kyoto \hfill}}
\def\titlepage{\FRONTPAGE\papers\ifPhysRev\PH@SR@V\fi
 \ifYUKAWA\null\vskip-1.70cm\YUKAWAmark\vskip0.6cm\fi
 \ifp@bblock\p@bblock \else\hrule height\z@ \rel@x \fi }

\def\schapter#1{\par \penalty-300 \vskip\chapterskip
 \spacecheck\chapterminspace
 \chapterreset \titlestyle{\ifcn@@\S\ \chapterlabel.~\fi #1}
 \nobreak\vskip\headskip \penalty 30000
 {\pr@tect\wlog{\string\chapter\space \chapterlabel}} }

\def\ssection#1{\par \ifnum\lastpenalty=30000\else
 \penalty-200\vskip\sectionskip \spacecheck\sectionminspace\fi
 \gl@bal\advance\sectionnumber by 1
 {\pr@tect
 \xdef\sectionlabel{\ifcn@@ \chapterlabel.\fi
 \the\sectionstyle{\the\sectionnumber}}%
 \wlog{\string\section\space \sectionlabel}}%
 \noindent {\S \caps\thinspace\sectionlabel.~~#1}\par
 \nobreak\vskip\headskip \penalty 30000 }


\papers

\def\lkakko{\vbox{\vskip0.065cm\hbox{(}\vskip-0.065cm}}
\def\rkakko{\vbox{\vskip0.065cm\hbox{)}\vskip-0.065cm}}
\def\YUKAWAHALL{\hbox to \hsize
 {\hfil \lkakko\twelvebf YUKAWA HALL\rkakko\hfil}}





\def\addeqno{\ifnum\equanumber<0 \global\advance\equanumber by -1
 \else \global\advance\equanumber by 1\fi}


\mathchardef\Lag="724C
\def\sqr#1#2{{\vcenter{\hrule height.#2pt
 \hbox{\vrule width.#2pt height#1pt \kern#1pt\vrule width.#2pt}
 \hrule height.#2pt}}}


\def\cref#1{\rlap,\attach{#1)}}
\def\ref#1{\attach{#1)}}



\newdimen\ex@
\ex@.2326ex
\def\boxed#1{\setbox\z@\hbox{$\displaystyle{#1}$}\hbox{\lower.4\ex@
 \hbox{\lower3\ex@\hbox{\lower\dp\z@\hbox{\vbox{\hrule height.4\ex@
 \hbox{\vrule width.4\ex@\hskip3\ex@\vbox{\vskip3\ex@\box\z@\vskip3\ex@}%
 \hskip3\ex@\vrule width.4\ex@}\hrule height.4\ex@}}}}}}
\def\txtboxed#1{\setbox\z@\hbox{{#1}}\hbox{\lower.4\ex@
 \hbox{\lower3\ex@\hbox{\lower\dp\z@\hbox{\vbox{\hrule height.4\ex@
 \hbox{\vrule width.4\ex@\hskip3\ex@\vbox{\vskip3\ex@\box\z@\vskip3\ex@}%
 \hskip3\ex@\vrule width.4\ex@}\hrule height.4\ex@}}}}}}
\newdimen\exx@
\exx@.1ex
\def\thinboxed#1{\setbox\z@\hbox{$\displaystyle{#1}$}\hbox{\lower.4\exx@
 \hbox{\lower3\exx@\hbox{\lower\dp\z@\hbox{\vbox{\hrule height.4\exx@
 \hbox{\vrule width.4\exx@\hskip3\exx@%
 \vbox{\vskip3\ex@\box\z@\vskip3\exx@}%
 \hskip3\exx@\vrule width.4\exx@}\hrule height.4\exx@}}}}}}

\chardef\fontD="1A

\catcode`@=12

\def\sskip{\vskip 5mm}
\def\dskip{\vskip 10mm}

\pubnum={YITP/K-1001}
\date={December 1992}
\titlepage
\parskip 0mm
\hsize=16cm
\vsize=23cm
\voffset=15mm

\foot{Talk given at the International Symposium on Rapidly Rotating
Nuclei 1992, Tokyo, 26-30 October 1992}

\vskip 15mm
\baselineskip 6mm
\noindent{\fourteenbf Chaotic Behavior in Warm Deformed Nuclei Induced
by \break
Residual Two-body Interactions}
\baselineskip 5mm

\sskip
\noindent M.Matsuo${}^{a}$, T.D\o ssing${}^{b}$, B.Herskind${}^{b}$,
S. Frauendorf${}^{\ c}$, E. Vigezzi${}^{d}$ and R. A. Broglia${}^{b,d,e}$
\sskip
\noindent ${}^a$ Yukawa Institute for Theoretical Physics, Kyoto University,
Kyoto 606-01, Japan
\sskip
\noindent ${}^b$ Niels Bohr Institute, University of Copenhagen,
DK-2100 Copenhagen \O , Denmark
\sskip
\noindent ${}^c$ Institut f\"ur Kern- und Hadronenphysik
FZ-Rossendorf, O-8054 Dresden, Germany
\sskip
\noindent ${}^d$ INFN Sezione di Milano, via Celoria 16, 20133 Milano, Italy
\sskip
\noindent ${}^e$ Dipartimento di Fisica dell'Universit\`a di Milano,
via Celoria 16, 20133 Milano, Italy
\dskip

\noindent{\twelvebf Abstract}

Band mixing calculations in rapidly rotating well-deformed nuclei are
presented, investigating the properties of energy levels and
rotational transitions as a function of excitation energy.
Substantial fragmentation of E2 transitions is found for
$E_x \gsim$ 800 keV above yrast, which represents the onset
of rotational damping. Above $E_x \approx $ 2 MeV, energy levels
and E2 strengths display fluctuations typical of quantum chaotic
systems, which are determined by the high multipole components
of the two-body residual interaction.

\dskip

\noindent{\bf 1. INTRODUCTION \hfil}
\sskip

It is a widely accepted
conjecture [1] that quantum chaotic systems exhibit
generic fluctuations governed by a random
matrix theory such as the Gaussian orthogonal ensemble (GOE).
The fluctuations refer not only to the energy levels
(nearest neighbor level spacing distribution, $\Delta_3$ statistics, etc)
but also to amplitudes of wave functions and strengths.
A typical example is provided by the strength fluctuations in
the widths of neutron resonances, which display a Porter-Thomas
distribution [2]. This shows that the nuclear system at an excitation
energy around $E_x \approx $ 8 MeV and angular momentum $\approx 1 \hbar$
is chaotic.
Recently, many attempts have been made to determine regions of
nuclear chaos varying parameters such as excitation
energy, spin, mass numbers, etc.
Rapidly rotating warm deformed nuclei, characterized by
$I \gsim 30 \hbar, E_x \lsim$ a few MeV above yrast line
in rare earth region are also investigated because
the rotational E2 transitions in well deformed nuclei
can be used as a measure of quantum chaos [3,4]. It is argued that
the rotational band structures correspond to regularity
while the rotational damping [5](disappearance of the band structure
associated with fragmentation of E2 strength) is
related to chaos
through complex mixing of rotational bands.
The rapidly rotating deformed nuclei at high excitation energy
($E_x \gsim$ several hundred keV above yrast) have high level density and
the detectors in experiments have a finite resolution.
Therefore, at the present stage of the development of experimental
techniques,
a statistical treatment of the  experimental data by a
fluctuation analysis of $E_\gamma - E_\gamma$
spectra [6] becomes more informative than discrete spectroscopy aiming
at identifying individual states.
If the observed E2 gamma-rays are emitted from a chaotic region,
the fluctuations of the $E_\gamma-E_\gamma$ spectra are governed
mainly by the fluctuations of E2 strengths while
the fluctuations of energy levels play a relatively minor role
due to their rigidity.
Hence, a study of
the strength fluctuation in stretched E2 transitions
will be useful in characterizing
rapidly rotating deformed nuclei at high excitation energies.

\dskip
\noindent{\bf 2. CRANKING MODEL WITH RESIDUAL TWO-BODY INTERACTION
\hfil}
\sskip

The cranking model is a simple and realistic model of
the rapidly rotating warm nuclei, especially
if many-particle many-hole excitations
as well as residual interactions among them are taken into account.
In this way thermal internal excitation
in the rotating nuclei is represented. In this work
we adopt the two-body surface delta interaction [7]
which will cause
mixing of the
unperturbed rotational bands with np-nh configurations.
The residual interaction
is diagonalized to give wave function contents of the mixed bands
(See ref.[8] for detail).
The procedure is similar to that followed
by S. \AA berg [4], who assumes however
constant two-body matrix elements with a random sign
irrespective of the cranked single-particle orbits involved.

\midinsert \vskip 9cm  \noindent
{\bf Figure 1.} The calculated rotational E2 strength distribution
as a function of the $\gamma$-ray energy $E_\gamma$ for four initial
levels with different excitation energies above yrast line
($E_x$ indicated in figure).
\endinsert

The following results are based on the
diagonalization for ${}^{168}$Yb at a
rotational frequency $\hbar\omega=0.5$ MeV, corresponding to
an average angular momentum $I \approx 40$. The diagonalization is
carried out for each parity and signature, with the lowest 1000
np-nh basis states yielding stable solutions for the lowest 300 levels
($E_x \lsim 2.3 $ MeV) above yrast.
The stretched  E2 strengths are calculated between levels
at the rotational frequency $\hbar\omega$ and those at
$\hbar\omega-2\hbar^2/\cal{J}$
$=0.47$ MeV, corresponding to a moment of inertia
$\cal{J}$$=67\hbar^2/$MeV.

Figure 1 illustrates how the rotational E2 transitions
change the structure with increasing excitation energy.
The E2 transition from the yrast
level shows a single peak in fig.1(a),
populating one final state belonging to
the same rotational band.
With increasing internal excitation energy of
the rotating nuclei,
the rotational transition branches into
several fragments
due to the band mixing caused by the surface delta interaction.
Actually,  fig.1(b) illustrates how the rotational transition
strength starts to become fragmented around an excitation energy
of $E_x \approx 800$ keV above yrast.
Further up in excitation energy (fig.1(c,d))
the rotational transitions exhibits not only fragmentation
but also a smooth profile as a function of
gamma-ray energy $E_\gamma$, from which a damping width (FWHM)
may be extracted.

\dskip
\noindent{\bf 3. THE ONSET OF ROTATIONAL DAMPING \hfil}
\sskip

\def\fig2{\vbox{\hbox{\hskip 9cm\vbox{
\hbox{{\bf Figure.2} \ The branching number aver-}
\hbox{aged over 100 keV intervals in the exci-}
\hbox{tation energy, and plotted as a function}
\hbox{of the energy above yrast. The onset of}
\hbox{rotational damping is marked by a bor-}
\hbox{derline $n_{branch}=2$.}}}}
}
\vskip 9.7cm \fig2
\vskip -12.28cm
\hangindent=-8cm \hangafter=-22
The structure change in the rotational E2 transitions
illustrated in fig.1 is
represented by characteristic excitation energies above yrast line.
As the excitation energy increases we first see the
onset of rotational damping
as fragmentation of E2 strength \break
from individual levels.
To quantitatively define the fragmentation of the E2
strength, we
introduce the branching number of level $i$:
$$
n_{branch}(i)=  \left (\sum_j  W_{ij}^2 \right )^{-1} \eqno{(1)}
$$
where $W_{ij}$ denotes the rotational transition probability
from level $i$ to $j$
($\propto$ E2 strength $S_{ij}$).
This expression is equal to the number of final states in the
case of equiprobable
branching $W_{ij}=1/n_{branch}$.
The branching number is shown as a function of
the excitation energy
in fig.2.
If the onset of
rotational damping is defined
by  a condition $n_{branch} > 2$,
the characteristic excitation energy $U_0$
for the onset of rotational damping is extracted from
fig. 2 as
$$
U_0 \approx 800 \  {\rm keV}\ \ . \eqno{(2)}
$$
The corresponding number of non-fragmented rotational bands, defined
by the condition $n_{branch} < 2$,
is 36. This number is in reasonable agreement with the recent result
of the fluctuation analysis [6] of the part of the
$E_\gamma-E_\gamma$ spectrum which displays rotational
energy correlations (the first ridge). The fluctuations of
that part of the spectrum are in accordance with 25 undamped bands.

\dskip
\noindent{\bf 4. CHAOTIC FLUCTUATIONS IN
ROTATIONAL DAMPING \hfil}
\sskip

As the excitation energy increases further,
strong band mixing is expected to take place.
If the mixing amplitudes become complex,
the E2 strengths in fragmented transitions
exhibit large fluctuations.
Figure 1(c) and (d) show sizable
fluctuation in the heights of
individual peaks.
If the strength fluctuations
of the E2 transitions obey the
Porter-Thomas distribution above a certain excitation energy $U_1$,
one may take this energy as the threshold for the realization
of quantum chaos in the system.

A quantitative analysis of the E2 strength fluctuation is
made by taking statistics (probability distribution)
of the E2 strengths.
The result
is shown in fig. 3. It should be noted that the {\it normalized}
strengths $s_{ij}=S_{ij}/\langle S_{ij} \rangle$ are analyzed here
instead of the  bare
\midinsert \vskip 8cm \noindent
{\bf Figure 3.} The probability distribution of the normalized
rotational E2 strengths $s_{ij}$ for gamma-ray
energies in the interval  $0.90 < E_\gamma < 1.05$ MeV.
The dashed curve represents
the Porter-Thomas distribution.
To display the dependence
on the excitation energy above yrast, the lowest 300 levels for each
parity and signature are grouped into four bins, containing
1-st to 50-th levels, 51- to 100-th levels, 101- to
200-th levels, and 201- to 300-th, and depicted in
(a),(b),(c),and (d), respectively. Excitation energies above yrast
of the bins are indicated in the figures.
\endinsert
\noindent E2 strength $S_{ij}$ in order to take
into account the smooth dependence  $\langle S_{ij} \rangle$
of the E2 strength on $E_\gamma$ and $E_x$.

Fluctuations in the energy levels are also
analyzed by means of the $\Delta_3$ statistics [2],
with the result shown in fig.5(a).
Figure 3 and fig.5(a) indicate that fluctuations in both the E2 strengths
and the energy levels reach the GOE limits
as the excitation energy above
yrast exceeds  about 2 MeV.
Thus the model predicts a characteristic excitation energy
of
$$
U_1 \approx 2 \ {\rm MeV} \ \ , \eqno{(3)}
$$
where the chaotic fluctuations set in.

\dskip
\noindent{\bf 5. MICROSCOPIC ORIGIN OF CHAOTIC BEHAVIOR \hfil}
\sskip

Let us investigate the origin of
the chaotic fluctuations in the rotational damping
in connection with the residual interactions among the unperturbed
rotational bands [9]. It is useful to decompose the residual
two-body interaction in terms of multipolarities of interacting
pair of nucleons, since much of our knowledge of the
residual two-body interaction has been acquired by studying
the nuclear response to probes  with specific multipolarity.
For example, the pairing plus quadrupole force
has been quite successful in accounting for many systematic
features of nuclear levels [10], but, it contains only
low multipolarities ($L=0,2$) by definition.
The surface delta interaction [7] (SDI),
on the other hand, contains all
the possible multipolarities as displayed in its expression:
\def\fig4{\vbox{\hbox{\hskip 8cm \vbox{
\hbox{\noindent{\bf Figure.4} Probability distribution of the off-}
\hbox{diagonal two-body matrix elements $V_{\alpha\beta\gamma\delta}$
of}
\hbox{the \ surface delta interaction (SDI) \ and the}
\hbox{pairing plus quadrupole force (P+QQ). The}
\hbox{dashed curve represents a Gaussian distribu-}
\hbox{tion which gives the same r.m.s. as SDI.}
}}}
}
\vskip 8.5cm \fig4
\vskip -11.5cm
\hangindent=-9cm \hangafter=-20
$$\eqalign{
V(1,2) &=  4 \pi V_o \delta(r_1 - R_o) \delta(r_2 - R_o) \ \ \ \cr
 &\times \sum_{\lambda, \mu} Y_{\lambda\mu}^* (\hat{r}_1)
Y_{\lambda\mu} (\hat{r}_2)\ \ ,   \ \ \  \  (4)
}$$
\hangindent=-9cm \hangafter=-10
\noindent where the sum of spherical harmonics denotes the delta function
of the angle between interacting pair.
Thus the two interactions differ in the
multipolarities they contain.

\hangindent=-9cm \hangafter=-11
A consequence of the difference between the two forces is seen
(fig.4) in the
probability
distribution of the two-body \ matrix \ elements
\ $V_{\alpha\beta\gamma\delta}$ \
for \  the \ \break
cranked-Nilsson s.p. orbits, denoted by $\alpha,\beta$,etc.
The strong peak at $V_{\alpha\beta\gamma\delta}=0$ for the
pairing plus quadrupole force indicates the presence of strong
selection rules while the distribution for the surface delta force
resembles much more a Gaussian shape,
\ which represents the

\vfil
\break
\noindent limit of
no selection rule. In other words, the higher multipole terms
contained in the SDI show considerably less selectivity
to specific orbits than the low multipole terms common to the SDI
as well as to
the pairing plus quadrupole interaction.

A comparison between the two forces is made in fig.5 both for
E2 strength  and  energy level fluctuations.
For fairness, a pairing plus quadrupole
force whose strength gives the same root-mean-square value
of the off-diagonal two-body matrix elements as that of
SDI (the selfconsistent values scaled by 1.5) is used here.
It is seen that
the pairing and quadrupole force does neither
give rise to the Porter-Thomas distribution
in E2 strength fluctuations nor
the GOE limit in the energy level fluctuations in the relevant energy region
($E_x \approx 2$ MeV), in contrast to the surface delta interaction,
which reaches to the chaotic limits (GOE).

Thus, it is a specific result of the present model of the rotational
damping  that
the high multipole components present in the residual
interaction are responsible for the mixing of rotational bands
and for
the chaotic fluctuations  (Porter-Thomas fluctuation in the E2
strength and the energy level fluctuations of GOE) in the energy
and spin region investigated.
This is because the
high multipole components of the two-body interaction
allow the unperturbed rotational bands to interact more
democratically irrespective of the participating single-particle
orbits than low multipole components.

\midinsert \vskip 10cm \noindent
{\bf Figure 5.}  Comparison between the surface delta interaction
and the paring plus \break
quadrupole force for the $\Delta_3$
statistics of energy level fluctuation (a) and for the
normalized E2 strength distribution (b). The three curves in (a),
marked by $a,b,c$, are the results for first to 100-th levels,
101- to 200-th levels, and 201- to 300-th levels for
each parity and signature, respectively.
The normalized strength distribution is shown for the
201- to 300-th levels.
\endinsert

\dskip
\noindent{\bf 6. CONCLUSIONS \hfil}
\sskip

It is shown that the band mixing caused by two-body residual interactions
brings about the rotational damping in rapidly rotating
warm deformed nuclei. With the surface delta interaction
acting among unperturbed rotational bands of np-nh
configurations, the rotational damping gradually sets in at around
excitation energy $U_0 \approx 800$ keV above yrast line,
at which the rotational E2 strength from individual level breaks into
more than two fragments. At higher excitation energy $E_x \gsim 2$ MeV,
the generic fluctuations proper of
quantum chaos (GOE) are found both in energy levels and in E2 strengths
(Porter-Thomas strength fluctuation).
{}From the comparison between the pairing plus quadrupole force
representing low multipole ($L=0,2$) components of the residual
two-body interaction and
the surface delta force including high multipole components as well,
it is found that
high multipole components in the two-body residual interaction
are responsible for the chaotic behavior of the rotational
damping.

\dskip
\line{\bf REFERENCES \hfill}
\sskip
\parindent=0mm

\def\refr{\par\hangindent=5mm\hangafter=-5\textindent}

\refr{1} O. Bohigas, M.J. Giannoni and C. Schmit, Phys. Rev. Lett.
{\bf 52}(1984),1.

\refr{2} As a review, T.A.Brody, J.Flores, J.B.French, P.A.Mello,
A.Pandy and S.S.M.Wong, Rev. Mod. Phys. {\bf 53}(1981), 385.

\refr{3} T.Guhr and H.A. Weidenm\"uller, Ann. Phys. {\bf 193}
(1989), 489.

\refr{4} S.\AA berg, Phys. Rev. Lett. {\bf 64}(1990), 3119;
Prog. Part. Nucl. Phys. Vol.28 (Pergamon 1992), p11.

\refr{5} B.Lauritzen, T.D\o ssing and R.A.Broglia, Nucl. Phys.
{\bf A457}(1986), 61.

\refr{6} B.Herskind, A.Bracco, R.A.Broglia, T.D\o ssing,
A.Ikeda, S.Leoni, J.Lisle, M.Matsuo, and E.Vigezzi, Phys. Rev. Lett.
{\bf 68}(1992), 3008. \hfil \break
B.Herskind, T.D\o ssing, S.Leoni, M.Matsuo, and E.Vigezzi,
Prog. Part. Nucl. Phys. Vol.28 (Pergamon 1992), p235.

\refr{7} I.M.Green and S.A. Mozkowski, Phys. Rev. {\bf 139}(1965),B790 ;
A. Faessler, Fort. der Physik {\bf 16}(1968),309 .

\refr{8} M. Matsuo, T. D\o ssing, B. Herskind and S. Frauendorf, Yukawa
Institute Kyoto preprint YITP/K-997 (1992).

\refr{9} M. Matsuo, T. D\o ssing, E. Vigezzi, and R.A. Broglia, Yukawa
Institute Kyoto preprint YITP/K-996 (1992).

\refr{10} A. Bohr and B.R. Mottelson, {\it Nuclear Structure.} vol. II
(Benjamin, 1975).

\bye